\RequirePackage{lineno}

\documentclass[preprint,tightenlines,twocolumn,superscriptaddress,amsmath,amssymb,nofootinbib]{revtex4-1}

\usepackage{graphicx} 
\usepackage{dcolumn} 
\usepackage{color}
\usepackage{epstopdf}
\usepackage[colorlinks]{hyperref}
\usepackage[capitalise]{cleveref}

\usepackage[separate-uncertainty = true, multi-part-units=single]{siunitx}

\usepackage{xspace}

\newcommand{\BSG}{\ensuremath{\bar{B}\to X_{s}\gamma}\xspace}
\newcommand{\BDG}{\ensuremath{\bar{B}\to X_{d}\gamma}\xspace}
\newcommand{\BSDG}{\ensuremath{\bar{B}\to X_{s+d}\gamma}\xspace}

\newcommand{\mxs}{\ensuremath{M_{X_{s}}}\xspace}

\newcommand{\BBbar}{\ensuremath{B\bar{B}}\xspace}
\newcommand{\epem}{\ensuremath{e^+e^-}\xspace}

\newcommand{\ecmg}{\ensuremath{E^{\text{*}}_{\gamma}}\xspace}
\newcommand{\pcml}{\ensuremath{p^{\text{*}}_{\ell}}\xspace}
\newcommand{\ups}{\ensuremath{\Upsilon(4S)}\xspace}

\newcommand{\MeV}{\ensuremath{\text{MeV}}\xspace}
\newcommand{\GeV}{\ensuremath{\text{GeV}}\xspace}
\newcommand{\ebg}{\ensuremath{E^{B}_{\gamma}}\xspace}
\newcommand{\mb}{\ensuremath{m_{b}}\xspace}
\newcommand{\mupi}{\ensuremath{\mu_{\pi}^{2}}\xspace}

\newcommand{\vub}{\ensuremath{|V_{ub}|}\xspace}
\newcommand{\vcb}{\ensuremath{|V_{cb}|}\xspace}

\newcommand{\nbb}{\ensuremath{N_{B\overline{B}}}\xspace}
\newcommand{\ethr}{\ensuremath{E^{B}_{\gamma}\geq E}\xspace}

\newcommand{\BFBSG}{\ensuremath{\mathcal{B}_{s\gamma}}}

\newcommand{\BFBSDG}{\ensuremath{\mathcal{B}_{s+d\gamma}}}

\DeclareSIUnit{\fb}{\femto\barn}
\DeclareSIUnit{\invfb}{\per\femto\barn}

\begin{document}

\preprint{\vbox{ \hbox{ }
 \hbox{ BELLE-CONF-1606}
}}

\title{\quad\\[1.0cm] Measurement of the inclusive \BSDG branching fraction, photon energy spectrum and HQE parameters}

\begin{abstract}
We report a measurement of the inclusive \BSDG and \BSG branching fractions using a data set of \num{772(11)e6} \BBbar pairs collected at the \ups resonance with the Belle detector at the KEKB asymmetric-energy \epem collider. Results are presented for photon energy thresholds between \numrange{1.7}{2.0}~\SI{}{GeV}. For the \SI{1.8}{\GeV} threshold we find  $\BFBSG = (3.01 \pm 0.10~\text{\small{(stat)}}  \pm 0.18~\text{\small{(syst)}} \pm 0.08~\text{\small{(model)}})\times 10^{-4}$. The Heavy Quark Expansion parameters that best fit the spectrum in the shape-function scheme are \mb=~\SI{4.626\pm 0.028}{\GeV} and \mupi=~\SI{0.301\pm 0.063}{\GeV^2}.

\end{abstract}

\date{\today}

\noaffiliation
\affiliation{Aligarh Muslim University, Aligarh 202002}
\affiliation{University of the Basque Country UPV/EHU, 48080 Bilbao}
\affiliation{Beihang University, Beijing 100191}
\affiliation{University of Bonn, 53115 Bonn}
\affiliation{Budker Institute of Nuclear Physics SB RAS, Novosibirsk 630090}
\affiliation{Faculty of Mathematics and Physics, Charles University, 121 16 Prague}
\affiliation{Chiba University, Chiba 263-8522}
\affiliation{Chonnam National University, Kwangju 660-701}
\affiliation{University of Cincinnati, Cincinnati, Ohio 45221}
\affiliation{Deutsches Elektronen--Synchrotron, 22607 Hamburg}
\affiliation{University of Florida, Gainesville, Florida 32611}
\affiliation{Department of Physics, Fu Jen Catholic University, Taipei 24205}
\affiliation{Justus-Liebig-Universit\"at Gie\ss{}en, 35392 Gie\ss{}en}
\affiliation{Gifu University, Gifu 501-1193}
\affiliation{II. Physikalisches Institut, Georg-August-Universit\"at G\"ottingen, 37073 G\"ottingen}
\affiliation{SOKENDAI (The Graduate University for Advanced Studies), Hayama 240-0193}
\affiliation{Gyeongsang National University, Chinju 660-701}
\affiliation{Hanyang University, Seoul 133-791}
\affiliation{University of Hawaii, Honolulu, Hawaii 96822}
\affiliation{High Energy Accelerator Research Organization (KEK), Tsukuba 305-0801}
\affiliation{J-PARC Branch, KEK Theory Center, High Energy Accelerator Research Organization (KEK), Tsukuba 305-0801}
\affiliation{Hiroshima Institute of Technology, Hiroshima 731-5193}
\affiliation{IKERBASQUE, Basque Foundation for Science, 48013 Bilbao}
\affiliation{University of Illinois at Urbana-Champaign, Urbana, Illinois 61801}
\affiliation{Indian Institute of Science Education and Research Mohali, SAS Nagar, 140306}
\affiliation{Indian Institute of Technology Bhubaneswar, Satya Nagar 751007}
\affiliation{Indian Institute of Technology Guwahati, Assam 781039}
\affiliation{Indian Institute of Technology Madras, Chennai 600036}
\affiliation{Indiana University, Bloomington, Indiana 47408}
\affiliation{Institute of High Energy Physics, Chinese Academy of Sciences, Beijing 100049}
\affiliation{Institute of High Energy Physics, Vienna 1050}
\affiliation{Institute for High Energy Physics, Protvino 142281}
\affiliation{Institute of Mathematical Sciences, Chennai 600113}
\affiliation{INFN - Sezione di Torino, 10125 Torino}
\affiliation{Advanced Science Research Center, Japan Atomic Energy Agency, Naka 319-1195}
\affiliation{J. Stefan Institute, 1000 Ljubljana}
\affiliation{Kanagawa University, Yokohama 221-8686}
\affiliation{Institut f\"ur Experimentelle Kernphysik, Karlsruher Institut f\"ur Technologie, 76131 Karlsruhe}
\affiliation{Kavli Institute for the Physics and Mathematics of the Universe (WPI), University of Tokyo, Kashiwa 277-8583}
\affiliation{Kennesaw State University, Kennesaw, Georgia 30144}
\affiliation{King Abdulaziz City for Science and Technology, Riyadh 11442}
\affiliation{Department of Physics, Faculty of Science, King Abdulaziz University, Jeddah 21589}
\affiliation{Korea Institute of Science and Technology Information, Daejeon 305-806}
\affiliation{Korea University, Seoul 136-713}
\affiliation{Kyoto University, Kyoto 606-8502}
\affiliation{Kyungpook National University, Daegu 702-701}
\affiliation{\'Ecole Polytechnique F\'ed\'erale de Lausanne (EPFL), Lausanne 1015}
\affiliation{P.N. Lebedev Physical Institute of the Russian Academy of Sciences, Moscow 119991}
\affiliation{Faculty of Mathematics and Physics, University of Ljubljana, 1000 Ljubljana}
\affiliation{Ludwig Maximilians University, 80539 Munich}
\affiliation{Luther College, Decorah, Iowa 52101}
\affiliation{University of Maribor, 2000 Maribor}
\affiliation{Max-Planck-Institut f\"ur Physik, 80805 M\"unchen}
\affiliation{School of Physics, University of Melbourne, Victoria 3010}
\affiliation{Middle East Technical University, 06531 Ankara}
\affiliation{University of Miyazaki, Miyazaki 889-2192}
\affiliation{Moscow Physical Engineering Institute, Moscow 115409}
\affiliation{Moscow Institute of Physics and Technology, Moscow Region 141700}
\affiliation{Graduate School of Science, Nagoya University, Nagoya 464-8602}
\affiliation{Kobayashi-Maskawa Institute, Nagoya University, Nagoya 464-8602}
\affiliation{Nara University of Education, Nara 630-8528}
\affiliation{Nara Women's University, Nara 630-8506}
\affiliation{National Central University, Chung-li 32054}
\affiliation{National United University, Miao Li 36003}
\affiliation{Department of Physics, National Taiwan University, Taipei 10617}
\affiliation{H. Niewodniczanski Institute of Nuclear Physics, Krakow 31-342}
\affiliation{Nippon Dental University, Niigata 951-8580}
\affiliation{Niigata University, Niigata 950-2181}
\affiliation{University of Nova Gorica, 5000 Nova Gorica}
\affiliation{Novosibirsk State University, Novosibirsk 630090}
\affiliation{Osaka City University, Osaka 558-8585}
\affiliation{Osaka University, Osaka 565-0871}
\affiliation{Pacific Northwest National Laboratory, Richland, Washington 99352}
\affiliation{Panjab University, Chandigarh 160014}
\affiliation{Peking University, Beijing 100871}
\affiliation{University of Pittsburgh, Pittsburgh, Pennsylvania 15260}
\affiliation{Punjab Agricultural University, Ludhiana 141004}
\affiliation{Research Center for Electron Photon Science, Tohoku University, Sendai 980-8578}
\affiliation{Research Center for Nuclear Physics, Osaka University, Osaka 567-0047}
\affiliation{Theoretical Research Division, Nishina Center, RIKEN, Saitama 351-0198}
\affiliation{RIKEN BNL Research Center, Upton, New York 11973}
\affiliation{Saga University, Saga 840-8502}
\affiliation{University of Science and Technology of China, Hefei 230026}
\affiliation{Seoul National University, Seoul 151-742}
\affiliation{Shinshu University, Nagano 390-8621}
\affiliation{Showa Pharmaceutical University, Tokyo 194-8543}
\affiliation{Soongsil University, Seoul 156-743}
\affiliation{University of South Carolina, Columbia, South Carolina 29208}
\affiliation{Stefan Meyer Institute for Subatomic Physics, Vienna 1090}
\affiliation{Sungkyunkwan University, Suwon 440-746}
\affiliation{School of Physics, University of Sydney, New South Wales 2006}
\affiliation{Department of Physics, Faculty of Science, University of Tabuk, Tabuk 71451}
\affiliation{Tata Institute of Fundamental Research, Mumbai 400005}
\affiliation{Excellence Cluster Universe, Technische Universit\"at M\"unchen, 85748 Garching}
\affiliation{Department of Physics, Technische Universit\"at M\"unchen, 85748 Garching}
\affiliation{Toho University, Funabashi 274-8510}
\affiliation{Tohoku Gakuin University, Tagajo 985-8537}
\affiliation{Department of Physics, Tohoku University, Sendai 980-8578}
\affiliation{Earthquake Research Institute, University of Tokyo, Tokyo 113-0032}
\affiliation{Department of Physics, University of Tokyo, Tokyo 113-0033}
\affiliation{Tokyo Institute of Technology, Tokyo 152-8550}
\affiliation{Tokyo Metropolitan University, Tokyo 192-0397}
\affiliation{Tokyo University of Agriculture and Technology, Tokyo 184-8588}
\affiliation{University of Torino, 10124 Torino}
\affiliation{Toyama National College of Maritime Technology, Toyama 933-0293}
\affiliation{Utkal University, Bhubaneswar 751004}
\affiliation{Virginia Polytechnic Institute and State University, Blacksburg, Virginia 24061}
\affiliation{Wayne State University, Detroit, Michigan 48202}
\affiliation{Yamagata University, Yamagata 990-8560}
\affiliation{Yonsei University, Seoul 120-749}
  \author{A.~Abdesselam}\affiliation{Department of Physics, Faculty of Science, University of Tabuk, Tabuk 71451} 
  \author{I.~Adachi}\affiliation{High Energy Accelerator Research Organization (KEK), Tsukuba 305-0801}\affiliation{SOKENDAI (The Graduate University for Advanced Studies), Hayama 240-0193} 
  \author{K.~Adamczyk}\affiliation{H. Niewodniczanski Institute of Nuclear Physics, Krakow 31-342} 
  \author{H.~Aihara}\affiliation{Department of Physics, University of Tokyo, Tokyo 113-0033} 
  \author{S.~Al~Said}\affiliation{Department of Physics, Faculty of Science, University of Tabuk, Tabuk 71451}\affiliation{Department of Physics, Faculty of Science, King Abdulaziz University, Jeddah 21589} 
  \author{K.~Arinstein}\affiliation{Budker Institute of Nuclear Physics SB RAS, Novosibirsk 630090}\affiliation{Novosibirsk State University, Novosibirsk 630090} 
  \author{Y.~Arita}\affiliation{Graduate School of Science, Nagoya University, Nagoya 464-8602} 
  \author{D.~M.~Asner}\affiliation{Pacific Northwest National Laboratory, Richland, Washington 99352} 
  \author{T.~Aso}\affiliation{Toyama National College of Maritime Technology, Toyama 933-0293} 
  \author{H.~Atmacan}\affiliation{Middle East Technical University, 06531 Ankara} 
% ...
\collaboration{The Belle Collaboration}



\maketitle

\renewcommand{\thefootnote}{\ifcase\value{footnote}\or $\dagger$ \or
$\ddagger$ \or $\S$ \or $\P$ \or ($\infty$)\fi}
\setcounter{footnote}{0}

The radiative transitions \BSG and \BDG proceed via electroweak loops at leading order in the Standard Model (SM), where a top quark and a charged weak boson are exchanged. These decays are sensitive to potential contributions from heavy non-SM particles. Theoretical calculations of the inclusive branching fraction are performed for the fully inclusive rate, equivalent to a photon energy threshold  of \SI{1.6}{\GeV}. Inclusive $B$ decays have the advantage that the decay rate, agrees with the decay rate of the free $b$ quark, up to small corrections due to the bound state. This makes it possible to have precise predictions. In contrast, calculations of exclusive decays suffer from large theoretical uncertainties, arising from the calculation of the $B$ form factors.

Measuring the full rate poses a challenge for experimentalists since background at low photon energies is orders of magnitude larger than the expected signal. The high precision of theoretical predictions has motivated a number of inclusive and semi-inclusive analyses. Inclusive analyses rely solely on finding the high-energy photon of the decay, and require a lot of work on reducing the background. Semi-inclusive analyses on the other hand, try to reconstruct dozens of the possible final states, facing different challenges such as missing modes and cross-feed. The current experimental world average~\cite{bib:hfag} is consistent with the most recent SM prediction of the \BSG branching fraction~\cite{bib:theo}\footnote{Observables for the decays \BSG, \BDG and \BSDG are identified with the subscripts $s\gamma$, $d\gamma$ and $s+d\gamma$, respectively.  We use natural units, e.g. $c = \hbar = 1$.}.
\begin{align}
&\BFBSG^{\text{SM}} = ( 3.36 \pm 0.23 )\times10^{-6},\\
&\BFBSG^{\text{exp}} = (3.43 \pm 0.21 \pm 0.07 )\times 10^{-6},
\end{align}
where the first uncertainty is statistical and the second systematic.  

The shape of the photon energy spectrum provides information about the Heavy Quark Expansion (HQE) parameters \mb and \mupi~\cite{bib:shape}.  These parameters are important in the determination of the magnitudes of the CKM matrix elements \vub and \vcb, when combined with measurements of semileptonic $B$ decays. Several theoretical descriptions of the spectrum are available, using different renormalization schemes such as the shape-function scheme~\cite{bib:BLNP}, kinetic scheme~\cite{bib:BBU} and the Kagan-Neubert model~\cite{bib:KN}.

In this letter we present an updated measurement of \BSG using the inclusive technique, which was performed on \num{657e6} \BBbar pairs~\cite{bib:tony}. We use the complete Belle data set of \SI{711}{\invfb} collected at the \ups energy (on-resonance sample), corresponding to \num{772(11)e6} \BBbar pairs. A \SI{90}{\invfb} sample to study light quark continuum background is taken at a \SI{60}{\MeV} lower energy (off-resonance sample). In addition to the larger sample size, this analysis uses multivariate tools to more effectively suppress continuum background, which greatly improves signal purity. We also present a novel determination of \mb and \mupi by folding the theoretical prediction of the shape-function scheme, and performing a fit to the measured spectrum. These parameters have usually been determined from using the spectral moments of semileptonic decays.

The Belle detector is located at the KEKB storage ring~\cite{bib:KEKB}. It is a large-solid-angle magnetic spectrometer consisting of a silicon vertex detector (SVD), a 50-layer central drift chamber (CDC), an array of aerogel threshold Cerenkov counters (ACC), a barrel-like arrangement of time-of-flight scintillation counters (TOF), and an electromagnetic calorimeter comprised of CsI(Tl) crystals (ECL) located inside a super-conducting solenoid coil that provided a \SI{1.5}{\tesla} magnetic field. An iron flux-return located outside of the coil is instrumented to detect $K_L^0$ mesons and to identify muons (KLM). The detector is described in detail elsewhere~\cite{bib:detector}.

In this analysis only the high energy photon from the signal decay is reconstructed, therefore we do not distinguish between the \BSG and \BDG contributions. The $X_d$ final state is suppressed by a factor $|V_{td}/V_{ts}|^2$ with respect to $X_s$, $|V_{td}/V_{ts}| = 0.216 \pm 0.011$~\cite{bib:pdg}. The \BSDG photon spectrum is obtained after subtracting high energy background photons from continuum and \BBbar events. Continuum background is subtracted using off-resonance data, and \BBbar background is estimated using Monte Carlo (MC) simulation, corrected using data control samples. We generate \num{2.6e6} \BSG MC events to optimize the selection criteria. This sample is composed of the resonant $K^*(892)$ meson for hadronic masses $\mxs<\SI{1.1}{\GeV}$ and an inclusive part that uses the Kagan-Neubert model for $\mxs\geq\SI{1.1}{\GeV}$. The inclusive $X_s$ particle is generated as a spin 1 $s\bar{d}$ or $s\bar{u}$ state and hadronized using JETSET~\cite{bib:jetset}. The low-\mxs and high-\mxs samples are mixed with a relative ratio consistent with the current world average~\cite{bib:hfag}.

Signal photon candidates are selected as connected clusters of ECL crystals with an energy in the center-of-mass frame (CM frame)\footnote{Quantities measured in the CM frame (the rest frame of the \ups) are starred, e.g. \ecmg to distinguish them from those measured in the $B$ meson rest frame, e.g. \ebg.} of $\SI{1.4}{\GeV}\leq \ecmg \leq \SI{4.0}{\GeV}$ in the angular region $\SI{32.2}{\degree}\leq \theta_{\gamma}\leq \SI{128.7}{\degree}$. The signal region is defined as $\SI{1.7}{\GeV}\leq \ecmg \leq \SI{2.8}{\GeV}$ and the sidebands below and above it are used to study \BBbar and continuum background, respectively. The ratio of the energy deposited in the $3\times3$ over $5\times5$ crystals around the seed crystal is required to be larger than $\SI{90}{\percent}$. Photons from $\pi^0(\eta)\to \gamma\gamma$ decays are vetoed using information from photon energy, polar angle and the reconstructed diphoton mass, as described in~\cite{bib:koppenburg}.

To suppress background from continuum events a lepton (electron or muon) consistent with a semileptonic decay of the other $B$ meson is reconstructed. The technique was also used to identify the flavor of the $B$ meson in~\cite{bib:acp}. The lepton is required to have a CM momentum \SI{1.10}{\GeV} $\leq \pcml\leq~$ \SI{2.25}{\GeV}. The selection on impact parameters with respect to the interaction point along the $z$ axis ($dz$) and perpendicular to it ($dr$) for the lepton track, are $|dz| \leq \SI{2}{cm}$ and $dr \leq \SI{0.5}{cm}$. We require at least one hit in the SVD and reconstruct tracks in the polar angle regions $\SI{18}{\degree}\leq \theta_{e}\leq \SI{150}{\degree}$ for electrons and  $\SI{25}{\degree}\leq \theta_{\mu}\leq \SI{145}{\degree}$ for muons.

To further suppress continuum background we trained boosted decision trees (BDT)~\cite{bib:tmva} using the following input variables: 11 modified Fox-Wolfram moments~\cite{bib:FW} constructed in 3 sets in which (1) we use all particles in the event, (2) we remove the signal photon and (3) we remove both signal photon and tag lepton; the distance between the photon to the closest extrapolated position on the ECL of a charged particle; the magnitude and direction of thrust, calculated from all charged and neutral particles in the event; the angle between the directions of the photon and tag lepton; the root mean square width of the photon cluster; the scalar sum of CM transverse momenta of all particles; the square of the missing four-momentum, calculated as the difference between the total beam energy and the momenta of all reconstructed particles. The BDT is trained using continuum and \BSG MC samples, and the selection criterion is chosen to maximize the expected statistical significance of the signal yield. We validate the performance of the BDT in a control sample of $\pi^0 \to \gamma \gamma$, which is obtained by requiring a large probability that the photon candidate originates from a $\pi^0$ decay and the difference between data and MC is interpreted as a systematic uncertainty.

After the selection we find \num{43008} events in the on-resonance and \num{702} in the off-resonance sample. The continuum yield must be scaled by a factor \num{7.509 \pm 0.196} to account for the difference in luminosities and $\epem\to q\bar{q}$ cross-sections of the on- and off-resonance samples. Continuum events are corrected to account for lower average particle energy and particle multiplicities. Continuum background composes about \SI{12}{\percent} of the selected on-resonance events, and the purity of \BSDG events is expected to be \SI{20}{\percent}.

The dominant \BBbar background sources are photons from $\pi^0$ and $\eta$ decays, contributing with \SI{49}{\percent} and \SI{8}{\percent} of the total yield respectively.  These events mimic the signal topology with one high energetic photon, and one of much lower energy. The normalization of the $\pi^0(\eta)$ background is performed by removing the $\pi^0(\eta)\to \gamma\gamma$ veto and then combining the prompt photon with any other photon in the event. For all combinations the diphoton invariant mass ($M_{\gamma\gamma}$) is calculated and a fit performed around the $\pi^0$ and $\eta$ masses to estimate their yields in data and MC. The fit is performed in 11 meson momentum bins in the range \numrange{1.4}{2.6}~\SI{}{\GeV} and the ratio of data to MC yields is used as a correction factor.

Photons from beam background make up roughly \SI{2}{\percent} of the total final sample. This contribution is determined from data using a sample of randomly triggered events overlaid with the MC. We assign a \SI{20}{\percent} uncertainty on its normalization. Clusters from neutral hadrons, particularly anti-neutrons or $K_L^0$, in the calorimeter could be misidentified as a photon cluster. This contribution is smaller than \SI{0.5}{\percent}. We correct the normalization of this component  since hadronic cluster shapes are not properly simulated in GEANT3~\cite{bib:geant}. While it is not possible to isolate a pure sample of neutrons, we estimate this component using anti-protons in $\Lambda \to p^- \pi^+$ decays. We found that the selection efficiency on the cluster shape criterion for anti-protons is underestimated by \SI{50}{\percent}. We scale the hadronic cluster contributions by a factor of 2 and assign a \SI{50}{\percent} uncertainty to account for any discrepancies between anti-proton and anti-neutron cluster shapes. Clusters from electrons without an associated track contribute less than \SI{1}{\percent} to the overall background, and we assign a \SI{20}{\percent} uncertainty on their yield.

The remaining \SI{6}{\percent} of the \BBbar background consists of photons from several sources: decays of $\omega$ and $\eta^{\prime}$ mesons, bremsstrahlung and others. None of the single contributions is significantly large, making it difficult to correct them individually. We scale this component to match data in the region $1.40 \leq \ecmg \leq \SI{1.55}{\GeV}$ with a factor of \num{1.30 \pm 0.15}, where the uncertainty is statistical.

\begin{figure}[htb]
\includegraphics[width=0.45\textwidth]{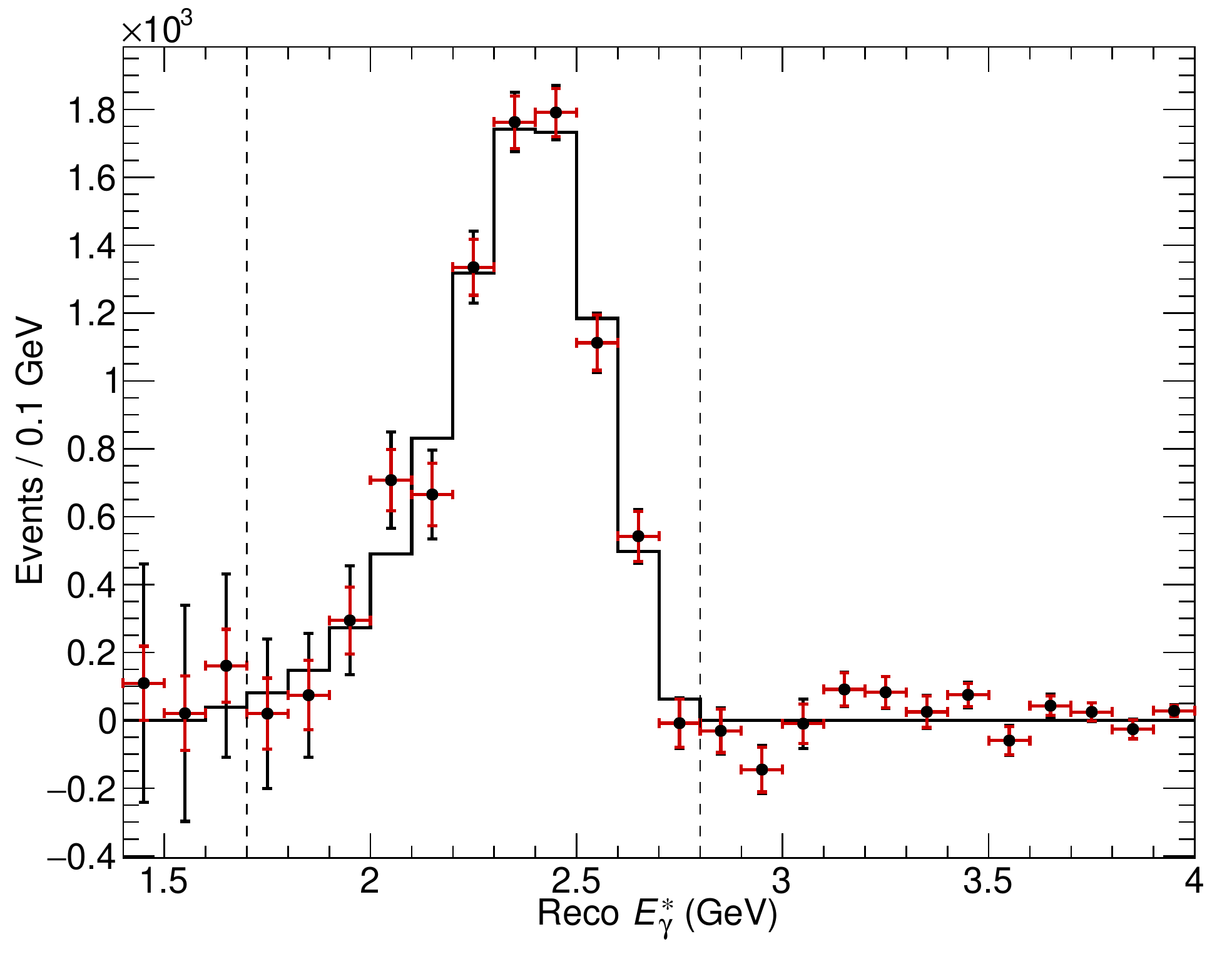}
\caption{Background-subtracted \BSDG photon energy spectrum. The internal (red) error bars represent statistical uncertainties, the outer (black) error bars represent total uncertainties. The histogram is the shape-function scheme spectrum with the best fit values for \mb and \mupi. The vertical lines show the measurement region.}
\label{fig:recocm}
\end{figure}

After subtracting all background contributions we find $8275 \pm 268~\text{\small{(stat)}} \pm 488~\text{\small{(syst)}}$ events in the region $1.8 \leq \ecmg \leq \SI{2.8}{\GeV}$. The background-subtracted spectrum is shown in~\cref{fig:recocm}, and is systematically limited in the low energy region where \BBbar background dominates.

Using the measured spectrum, we determine the HQE parameters \mb and \mupi using the shape-function scheme. The theoretical calculation is carried out in the $B$ rest frame, whereas the reported spectrum is measured in the CM frame. The calculation assumes that any resonant structure is sufficiently broadened by the experimental resolution, and the spectrum can be fully described inclusively. This is reasonable as the calorimeter resolution and  Doppler broadening  would together make it impossible to resolve any resonant structure. We use MC simulation to include these effects in the theoretical expectation. After this we apply selection efficiency effects, and the theoretical and measured spectra are compared. We perform a chi-squared fit in which \mb and \mupi are free parameters and use the full experimental covariance matrix of the background-subtracted spectrum. The fit is performed in the photon energy region $1.8 \leq \ecmg \leq \SI{2.8}{\GeV}$, and we find \mb=~\SI{4.626\pm 0.028}{\GeV} and \mupi=~\SI{0.301\pm 0.063}{\GeV^2} with a correlation of $\rho = -0.701$.

We also report the inclusive \BSDG branching fraction for various energy thresholds.  
\begin{equation}
\BFBSDG^{\ethr} = \frac{1}{\varepsilon_{\text{rec}}}\cdot \frac{\alpha^{\ethr}}{2\nbb~\varepsilon_{\text{eff}}}N^{\text{\ethr}}.
\label{eq:bf}
\end{equation}
Here $N^{\ethr}$ is the integral of the spectrum for a given threshold, $\varepsilon_{\text{eff}}$  is the selection efficiency, \nbb is the total number of \BBbar pairs and $\varepsilon_{\text{rec}}$ is the efficiency for a signal photon to be reconstructed in the calorimeter. The factor $\alpha^{\ethr}$ transforms the measurement at a given threshold energy from the CM frame to the $B$ meson rest frame. Both the selection efficiency and $\alpha^{\ethr}$ are model dependent. To calculate them, we fit our spectrum using the Kagan-Neubert, kinetic and shape-function models, determine then the quantities, and take the average among them for the central value. The model uncertainty is assigned as the largest deviation between the average and the best fit values $\pm1\sigma$. The average selection efficiency is $\varepsilon_{\text{eff}} = \SI{2.45}{\percent}$ at the \SI{1.8}{\GeV} threshold, the efficiency $\varepsilon_{\text{rec}}$ takes a value of 0.712. To obtain the \BSG branching fraction we divide the \BSDG  result by a factor of $1+|V_{td}/V_{ts}|^2$. The underlying assumption is that inclusive shape for both final states is identical, which is reasonable given that both are two-body decays, and the detector and Doppler effect would effectively broaden the contributions from resonances in the decay.

The results for the inclusive \BSDG and \BSG branching fractions for energy thresholds between \numrange{1.7}{2.0}~\SI{}{\GeV} are summarized in~\cref{table:bf} and the corresponding correlations in~\cref{table:correlation}. The largest systematic uncertainty arises from the corrections and uncertainties associated with the subtraction of the \BBbar background, which is \SI{5.2}{\percent} at the \SI{1.8}{\GeV} threshold. Additional uncertainties arise from the continuum background subtraction, \SI{1.3}{\percent}, the number of \BBbar pairs,  \SI{1.4}{\percent}, and the BDT selection, \SI{1.4}{\percent}. This last systematic uncertainty accounts for possible BDT mis-modeling in the \BBbar MC and is studied in a $\pi^0$ control sample.
\begin{table*}[htb]
  \begin{tabular}{l|cccc}
  Threshold & Selection eff. (\SI{}{\percent}) & $\alpha^{\ethr}$ & $\BFBSDG$ & $\BFBSG$ \\
   \hline
  \SI{1.7}{\GeV} & $2.392 \pm 0.070$ & $1.0135 \pm 0.0024$ & $ 3.20 \pm 0.11 \pm 0.25 \pm 0.10 $ & ~$ 3.06 \pm 0.11 \pm 0.24 \pm 0.09 $\\
  \SI{1.8}{\GeV} & $2.442 \pm 0.059$ & $1.0216 \pm 0.0031$ & $ 3.15 \pm 0.10 \pm 0.19 \pm 0.08 $ & ~$ 3.01 \pm 0.10 \pm 0.18 \pm 0.08 $\\
  \SI{1.9}{\GeV} & $2.508 \pm 0.055$ & $1.0334 \pm 0.0039$￼ & $ 3.07 \pm 0.09 \pm 0.15 \pm 0.07 $ & ~$ 2.94 \pm 0.09 \pm 0.14 \pm 0.07 $\\
  \SI{2.0}{\GeV} & $2.595 \pm 0.045$ & $1.0526 \pm 0.0046$￼ & $ 2.92 \pm 0.08 \pm 0.11 \pm 0.05 $ & ~$ 2.79 \pm 0.08 \pm 0.11 \pm 0.05 $\\
  \end{tabular}
  \caption{Inclusive \BSDG and \BSG branching fractions for different energy thresholds up to \SI{2.8}{\GeV}, in units of $10^{-4}$. The uncertainties are statistical, systematic and from the modeling.}
  \label{table:bf}
\end{table*}

\begin{table}[htb]
\begin{tabular}{l|cccc}
 & \SI{1.7}{\GeV} & \SI{1.8}{\GeV} & \SI{1.9}{\GeV} & \SI{2.0}{\GeV}\\
     \hline
\SI{1.7}{\GeV} & 1.00 & 0.92 & 0.83 & 0.72 \\
\SI{1.8}{\GeV} & & 1.00 & 0.91 & 0.81 \\
\SI{1.9}{\GeV} & & & 1.00 & 0.90  \\
\SI{2.0}{\GeV} & & & & 1.00  \\
    \end{tabular}
    \caption{Correlation between \BSG branching fraction measured for different thresholds.}
    \label{table:correlation}
\end{table}

\begin{figure}[htb]
\includegraphics[width=0.45\textwidth]{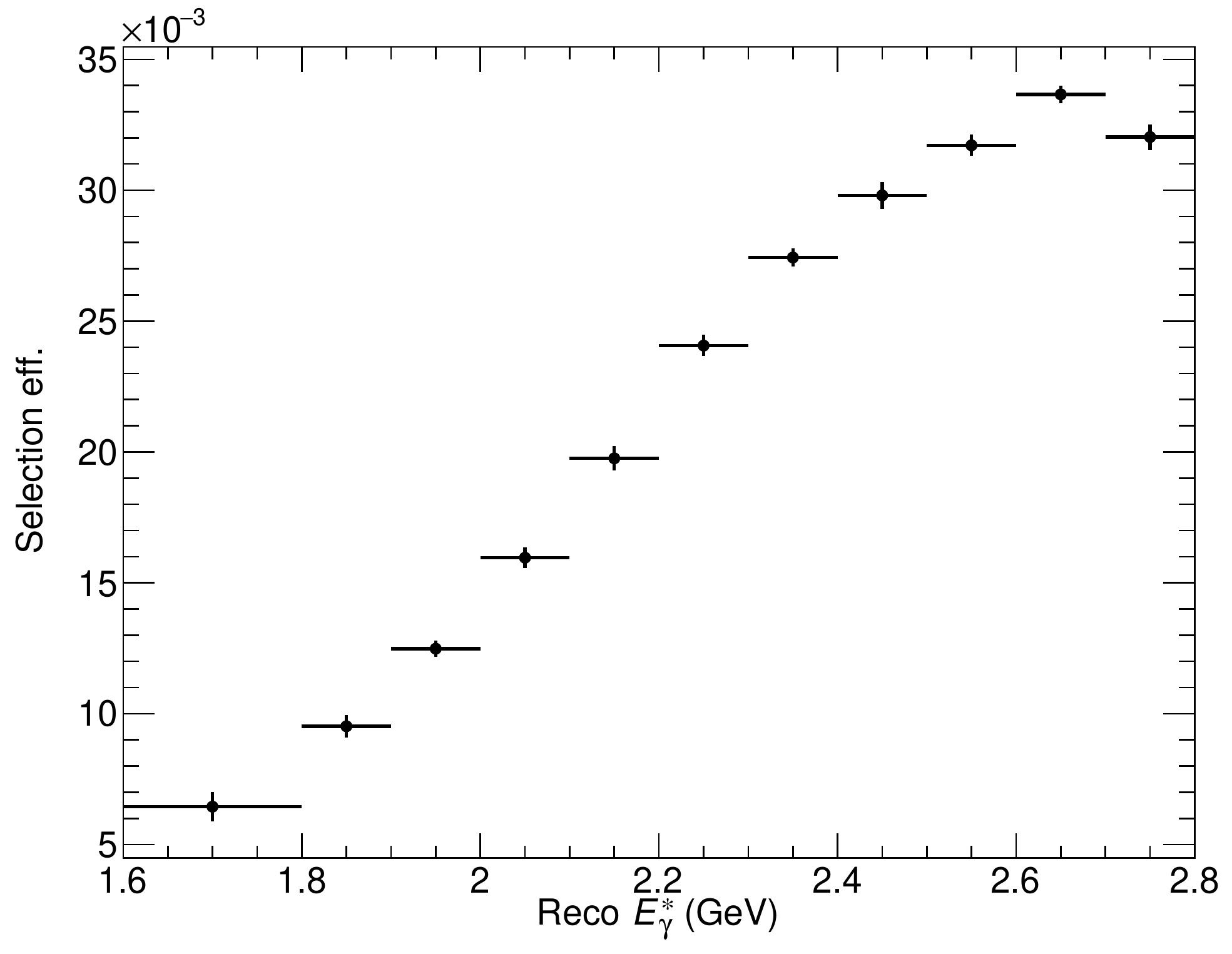}
\caption{Selection efficiency determined from the average among three models for the \BSG spectrum (see text). The uncertainty is determined by the largest deviation between the average and the 1$\sigma$ errors at the best fit values for the three models.}
\label{fig:selectioneff}
\end{figure}

In order to measure the partial branching fractions and the moments of the spectrum, we must correct it according to the selection efficiency in each \ecmg bin, unfold the resolution and migration effects caused by the finite energy resolution of the ECL and correct for the reconstruction efficiency $\varepsilon_{\text{rec}}$. The bin-by-bin selection efficiency is summarized in~\cref{fig:selectioneff}. The unfolding procedure is based on the Singular Value Decomposition algorithm~\cite{bib:svd}. The covariance matrix has large correlations and systematic uncertainties at low photon energy. This causes the default algorithm to systematically underestimate the uncertainties after unfolding and regularization. The problem was caused by the rescaling of equations performed in the algorithm~\cite{bib:kerstin}, we remove it by skipping the step of equation 34 of~\cite{bib:svd}. The systematic uncertainty related to the unfolding procedure is determined using an ensemble of spectra from the three available models, it is much smaller than statistical uncertainties and uncertainties from \BBbar background suppression, ranging from $\SI{3.5}{\percent}$ for the $1.8\leq \ebg \leq \SI{1.9}{\GeV}$ bin, to $\sim\SI{0.1}{\percent}$ above \SI{2.2}{\GeV}. We only report unfolded spectra in the CM frame, as unfolding to the $B$ rest frame introduces high model dependence, the transformation of quantities calculated in the $B$ rest frame into the CM frame is simple from the theory side and has been discussed in~\cite{bib:KN}.

Finally, we also report the first and second spectral moments, which correspond to the average energy of the spectrum and the variance and contain information about the HQE parameters. They are obtained for different energy thresholds. The partial branching fractions are summarized in~\cref{table:pbf} and the spectral moments in~\cref{table:moments}. 
\begin{table}[htb]
\begin{tabular}{l|ccccc}
    \ecmg bin (\SI{}{\GeV}) & Total & PBF & Stat & Syst & Model \\
    \hline
    1.6-1.8 & 12.3 & 91.1 & 29.9 & 86.1 & 2.3 \\
    1.8-1.9 & 11.6 & 51.6 & 15.7 & 49.1 & 1.3 \\
    1.9-2.0 & 16.7 & 28.1 & 10.8 & 26.0 & 1.0 \\
    2.0-2.1 & 24.2 & 14.9 & 9.6 & 11.3 & 1.1 \\
    2.1-2.2 & 34.7 & 11.8 & 8.1 & 8.4 & 1.6 \\
    2.2-2.3 & 47.6 & 9.4 & 7.1 & 5.9 & 1.9 \\
    2.3-2.4 & 61.1 & 7.6 & 6.4 & 3.5 & 1.8 \\
    2.4-2.5 & 63.1 & 6.6 & 5.6 & 2.8 & 2.3 \\
    2.5-2.6 & 43.7 & 5.2 & 4.7 & 1.9 & 1.1 \\
    2.6-2.7 & 20.1 & 4.4 & 4.1 & 1.6 & 0.5 \\
    2.7-2.8 & 1.9 & 0.6 & 0.5 & 0.2 & 0.1 \\
    \end{tabular}
    \caption{Partial branching fractions of the \BSDG spectrum and uncertainties, in units of $10^{-6}$.}
    \label{table:pbf}
\end{table}

\begin{table*}[htb]
\begin{tabular}{l|cc}
    Threshold & Mean~\SI{}{(\GeV)} & Variance$\times 10^2$~\SI{}{(\GeV\squared)} \\
    \hline
    \SI{1.8}{\GeV} & $ 2.320 \pm 0.034 \pm 0.105 \pm 0.003 $ & ~$ 4.258 \pm 1.123 \pm 3.451 \pm 0.108 $ \\
    \SI{1.9}{\GeV} & $ 2.338 \pm 0.022 \pm 0.041 \pm 0.003 $ & ~$ 3.563 \pm 0.533 \pm 1.047 \pm 0.065 $ \\
    \SI{2.0}{\GeV} & $ 2.360 \pm 0.015 \pm 0.017 \pm 0.003 $ & ~$ 2.869 \pm 0.292 \pm 0.282 \pm 0.047 $ \\
    \end{tabular}
    \caption{Mean and variance of the \BSDG spectrum in the CM frame. The uncertainties are statistical, systematic and from model dependence.}
    \label{table:moments}
\end{table*}

Using our ensemble of theoretical descriptions of the spectrum, we determine an extrapolation factor to translate the measured branching fraction from the \SI{1.8}{\GeV} threshold to \SI{1.6}{\GeV}. The central values and uncertainties are determined in the same fashion as the selection efficiency and $\alpha^{\ethr}$. We find a factor $1.0369 \pm 0.0139$ and extract $\BFBSG^{\ebg>1.6} = (3.12 \pm 0.10~\text{\small{(stat)}} \pm 0.19~\text{\small{(syst)}} \pm 0.08~\text{\small{(model)}} \pm 0.04~\text{\small{(extrap)}})\times 10^{-4}$, which is in agreement with the SM prediction, as well as previous experimental measurements. The extrapolation factors used by HFAG are determined from fits to \BSG and $B\to X_c \ell \nu$ moments~\cite{bib:hfagfactors}. Similarly to our determination, the factors are obtained averaging the values from the three theoretical models we have also uses, the factor for the \SI{1.8}{\GeV} threshold is perfectly compatible to our determination. We use the extrapolated inclusive branching fraction $\BFBSG^{\ebg>1.6}$ to find a lower bound on the mass of a charged Higgs boson in the framework of the type-II Two-Higgs-Double-Model (2HDM-II). Using the procedure described in~\cite{bib:ckmfitter}, we exclude $M_{H^{\pm}}$ smaller than $\SI{580}{\GeV}$ with a \SI{95}{\percent} confidence level.

\vspace{1cm}
We thank the KEKB group for excellent operation of the accelerator; the KEK cryogenics group for efficient solenoid operations; and the KEK computer group, the NII, and PNNL/EMSL for valuable computing and SINET4 network support. We acknowledge support from MEXT, JSPS and Nagoya's TLPRC (Japan); ARC (Australia); FWF (Austria); NSFC and CCEPP (China); MSMT (Czechia); CZF, DFG, EXC153, and VS (Germany); DST (India); INFN (Italy); MOE, MSIP, NRF, BK21Plus, WCU and RSRI  (Korea); MNiSW and NCN (Poland); MES and RFAAE (Russia); ARRS (Slovenia); IKERBASQUE and UPV/EHU (Spain); SNSF (Switzerland); MOE and MOST (Taiwan); and DOE and NSF (USA).

\end{document}